\documentclass[3p,twocolumn]{elsarticle}      

\usepackage{graphicx}

\begin{document}

\title{Optimized t-expansion method for the Rabi Hamiltonian}

\author{Igor Trav\v{e}nec \corref{c1}}
\author{Ladislav \v{S}amaj}
\address{Institute of Physics, Slovak Academy of Sciences, \\
D\'ubravsk\'a cesta 9, 845 11 Bratislava, Slovakia}
\cortext[c1]{Corresponding author, e-mail: fyzitrav@savba.sk}

\begin{abstract}
A polemic arose recently about the applicability of the $t$-expansion 
method to the calculation of the ground state energy $E_0$ of 
the Rabi model.
For specific choices of the trial function and very large number of
involved connected moments, the $t$-expansion results are rather poor
and exhibit considerable oscillations.
In this letter, we formulate the $t$-expansion method for trial functions 
containing two free parameters which capture two exactly solvable limits 
of the Rabi Hamiltonian.
At each order of the $t$-series, $E_0$ is assumed to be stationary with 
respect to the free parameters.
A high accuracy of $E_0$ estimates is achieved for small numbers (5 or 6)
of involved connected moments, the relative error being smaller than $10^{-4}$ 
($0.01$\%) within the whole parameter space of the Rabi Hamiltonian.
A special symmetrization of the trial function enables us to calculate also 
the first excited energy $E_1$, with the relative error smaller than $10^{-2}$
(1\%). 
\end{abstract}


\maketitle

\noindent {\bf Introduction:}
This paper is about the $t$-expansion method of the calculation of
low-lying energy spectrum for quantum Hamiltonian systems.
The application of the method to the Rabi model evoked some doubts about 
its reliability \cite{Fess02b,Amo}.
In this letter, we propose such treatment of $t$-expansion series which 
provides, in low approximation orders, extraordinarily accurate estimates 
of the ground state energy in the whole range of model's parameters.  
First we explain the $t$-expansion method, then summarize the variational 
approaches to the Rabi Hamiltonian, propose a stationarity treatment
of the $t$-expansion with two-parameter trial functions and finally
present accurate numerical results. 
\medskip

\noindent {\bf t-expansion:}
The $t$-expansion technique, originated by Horn and Weinstein \cite{HW}, 
is a ``series extension'' of the variational method.
It is based on the following theorem. 
For any trial function $\vert\psi\rangle$, which has a non-zero overlap with 
the exact ground state of a Hamiltonian $\hat{H}$, the function
\begin{equation} \label{e1}
E(t) = \frac{ \langle\psi\vert\hat{H} e^{-t\hat{H}}\vert\psi\rangle}{
\langle\psi \vert e^{-t\hat{H}} \vert \psi\rangle} 
= \sum_{m=0}^\infty \frac{(-t)^m}{m!}I_{m+1}
\end{equation}
monotonously decays in $t$, approaching the ground-state 
energy $E_0$ at asymptotically large $t$: $\lim_{t\to\infty}E(t)=E_0$.
The coefficients of the small-$t$ expansion $\{ I_m \}$ are known as 
the connected moments.
They can be expressed recursively in terms of the standard moments 
$\mu_m=\langle\psi\vert\hat{H}^m\vert\psi\rangle$ as
\begin{equation} \label{e2}
I_m = \mu_m - \sum_{k=0}^{m-2} {m-1\choose k} I_{k+1}\mu_{m-k-1}
\end{equation}
for $m\ge 2$ and $I_1=\mu_1$.
The $t=0$ estimate $E(0)=I_1$ is the variational value, representing 
a rigorous upper bound for the ground state energy.

Usually only a limited number of connected moments can be evaluated.
Since we are interested in the $t\to\infty$ behavior of $E(t)$, one needs some 
extrapolation from the small-$t$ series to large $t$. 
From among various schemes \cite{Cio,Stu,CSM,Fess05} we choose 
the following ones.
The widely used Connected moments expansion (CMX) \cite{Cio}
considers $E(t)$ to be a sum of exponentials. 
The estimate of $E_0$, available only for odd number of connected moments
$m=2k+1$ $(k=0,1,\ldots)$, is given by 
\begin{equation} \label{e3}
E_0^{(m)} = I_1 -{\bf X}_k {\bf T}_k^{-1} {\bf X}_k^{\rm T} ,
\end{equation}
where the vector ${\bf X}_k=(I_2, ...,I_{k+1})$ and the matrix ${\bf T}_k$ 
has elements $(T_k)_{ij}=I_{i+j+1}$, $(i,j=1,\ldots,k)$. 
In particular, we have $E_0^{(1)}=I_1$, $E_0^{(3)}=I_1-I_2^2/I_3$, etc.
The second scheme, known in the literature \cite{Fess02,Fess05b,Fess08} as 
the Canonical sequence method (CSM) \cite{CSM}, corresponds to 
a polynomial ``deformation'' of one exponential.
The method is formulated in the inverse format, using the function $t(E)$ 
instead of $E(t)$; the series expansion of $t(E)$ around $E=I_1$ is deducible 
from (\ref{e1}) \cite{Stu}. 
The estimate of the ground state energy, which involves $m$ connected 
moments ($m$ may be even or odd), reads 
\begin{equation} \label{e4}
E_0^{(m)}=I_1+(m-2)\frac{{\rm d}_E^{(m-2)}t(I_1)}{{\rm d}_E^{(m-1)}t(I_1)},
\end{equation}
where ${\rm d}_E^{(k)}t(I_1)$ means the $k$-th derivative of $t(E)$ at $E=I_1$.
We have the same $E_0^{(1)}$ and $E_0^{(3)}$ as in CMX,
$E_0^{(4)}=I_1+2I_2^2I_3/(I_2I_4-3I_3^2)$, etc. 
\medskip

\noindent {\bf Rabi Hamiltonian:}
The Rabi model \cite{Rabi} describes the interaction between a bosonic mode 
with energy $\omega\ge 0$ and a two-level atom with the gap $\omega_0$. 
Its Hamiltonian is 
\begin{equation} \label{e6}
\hat{H} = \frac{1}{2}\omega_0\sigma^z+\omega b^\dagger b
+g(\sigma^+ +\sigma^-)(b^\dagger+b),
\end{equation}
where $g$ is the interaction constant, $\sigma^z$, $\sigma^{\pm}$ are 
the Pauli matrices, $b^\dagger$ and $b$ are boson creation and 
annihilation operators, respectively.

There exist two exactly solvable cases of the Rabi Hamiltonian. 
For $g=0$ the system decouples and the ground-state wavefunction
is the tensor product
\begin{equation} \label{b1}
\vert\psi\rangle = {0\choose 1} \vert 0\rangle ,
\end{equation} 
as the atom stays at the bottom level and the boson is in his lowest mode 
as well. 
The energy of this state is $E_0=-\omega_0/2$. 
The other case is $\omega_0=0$, when the two atomic levels merge to 
a degenerate one \cite{Bish}. 
The exact (two-fold degenerate) ground state 
\begin{equation} \label{e7}
\vert \psi(x,y)\rangle = \frac{1}{\sqrt{y^2+1}}
{y\choose 1} \vert x\rangle
\end{equation}
is the product of the eigenfunction of $(\sigma^+ +\sigma^-)=2\sigma^x$ and 
the coherent boson state
\begin{equation} \label{e8}
\vert x\rangle = \exp \left( - \frac{x^2}{2} + x b^\dagger \right) 
\vert 0\rangle .
\end{equation}
One ground state is specified by the parameters $\{ x = 2g/\omega, y=-1 \}$, 
the conjugate one by the oppositely signed $\{ x=-2g/\omega, y=1 \}$. 
The ground state energy is $E_0 = - 4g^2/\omega$. 

The (trial) function (\ref{e7}) is intentionally written so that both $x$ and 
$y$ can serve as variational parameters for an arbitrary $\omega_0\ne 0$, 
as was done in the standard variational method \cite{Bish,Qin}. 
The optimized values of the parameters $x_{\rm opt}$ and $y_{\rm opt}$ are 
determined by minimizing $E_0^{(1)}(x,y) = I_1(x,y)$: 
\begin{equation} \label{b2}
\frac{\partial}{\partial x} E_0^{(1)} = 
\frac{\partial}{\partial y} E_0^{(1)} = 0 .
\end{equation} 
The interpolation between the two exact solutions,
$\{ x=0, y=0\}$ at $g=0$ (\ref{b1}) and the branch $\{ x = 2g/\omega, y=-1 \}$
at $\omega_0=0$, is provided by the unique solution 
$\{ x_{\rm opt},y_{\rm opt} \}$ with components restricted to the intervals
\begin{equation} \label{e9}
0\le x_{\rm opt}\le \frac{2g}{\omega}, \qquad -1\le y_{\rm opt}\le 0.
\end{equation}

Bishop et al. \cite{Bish} pointed out a conserved parity of 
the Rabi Hamiltonian, associated with the sign reversal transformation 
$\{x, y\} \to \{-x,\ -y\}$.
The trial function (\ref{e7}), which does not possess this symmetry,
will be referred to as non-symmetrized. 
Two symmetrized versions of (\ref{e7}) were proposed:
\begin{eqnarray} \label{e10}
\vert \psi^{(p,n)}(x,y)\rangle = c_{\pm}(x,y) 
\left[ \vert x\rangle\pm \vert -x\rangle \right] {{0}\choose{1}} \nonumber \\ 
+ y\, c_{\mp}(x,y) \left[ \vert x\rangle\mp \vert -x\rangle \right]
{1\choose 0} ,
\end{eqnarray}
where the normalization constants are
\begin{equation} \label{e11}
c_{\pm}(x,y) = \frac{1}{\sqrt{y^2+1}}\frac{1}{\sqrt{2\pm 2e^{-2x^2}}} .
\end{equation}
$(p)$ and $(n)$ stand for the positive and negative parity, respectively. 
It can be shown that $\vert\psi^{(p)}(x,y)\rangle$ and 
$\vert\psi^{(n)}(x,y)\rangle$ are orthogonal to one another.
Within the variational approach \cite{Bish}, $\vert\psi^{(p)}(x,y)\rangle$ 
implies the ground state energy $E_0$ of parity $(p)$
The application of the $(n)$-symmetrized trial function 
$\vert\psi^{(n)}(x,y)\rangle$ projects the ground state away 
and, consequently, implies the first excited energy $E_1$ of parity $(n)$.

To calculate moments of the Rabi Hamiltonian with an arbitrary one of 
the three trial functions $\vert\psi(x,y)\rangle$ or 
$\vert\psi^{(p,n)}(x,y)\rangle$, we apply the commutator 
$b b^{\dagger}=1+b^{\dagger} b$ and the useful formula 
\begin{equation} \label{e12}
\langle x_1\vert (b^{\dagger})^k b^l \vert x_2\rangle =
x_1^k x_2^l e^{-(x_1-x_2)^2/2}
\end{equation}
($k,l=0, 1, 2,\ldots$) valid for each of four possibilities 
$x_1=\pm x$, $x_2=\pm x$. 
The moments with the non-symmetrized trial function 
$\mu_m = \langle\psi(x,y)\vert\hat{H}^m\vert\psi(x,y)\rangle$ are found to be
\begin{eqnarray} \label{e13}
\mu_1 = \omega x^2 +\frac{8 g x y}
{y^2+1}+\frac{\omega_0}{2}\frac{y^2-1}{y^2+1}, \phantom{aaaaa}\nonumber \\
\mu_2 = \frac{\omega_0^2}{4}+4g^2(1+4x^2)+\omega^2(x^2+x^4)\nonumber \\ 
+\frac{8g\omega x y(1+2x^2)+\omega_0\omega x^2(y^2-1)}{y^2+1},
\end{eqnarray}
etc. 
The moments with the $(p)$ and $(n)$ symmetrized trial functions are
obtained in the form
\begin{eqnarray}
\mu_1^{(p,n)} & = & \frac{\omega_0}{2}\frac{y^2-1}{y^2+1}+\frac{8gxy}
{(1+y^2)\sqrt{1-e^{-4x^2}}}\nonumber \\ 
& & +\frac{\omega x^2}{y^2+1} \left[ y^2(\coth{x^2})^{\pm 1} \right. 
\nonumber \\ & & \left. \qquad \qquad 
+(\tanh{x^2})^{\pm 1} \right], \nonumber \\
\mu_2^{(p,n)} & = & \frac{\omega_0^2}{4} + \omega^2 x^4+4g^2(1+2x^2)
\nonumber \\
& & +\frac{8g\omega x y(1+2x^2)}{(y^2+1)\sqrt{1-e^{-4x^2}}}\nonumber \\
& & +\frac{\omega_0\omega x^2}{y^2+1} 
\left[ y^2(\coth{x^2})^{\pm 1} \right. \nonumber \\
& & \left. \qquad \qquad -(\tanh{x^2})^{\pm 1}\right]  \nonumber \\
& & +\frac{(8g^2+\omega^2)x^2}{y^2+1} \left[ y^2(\coth{x^2})^{\pm 1}
\right. \nonumber \\ & & \left. \qquad \qquad + (\tanh{x^2})^{\pm 1} \right] ,
\label{e14}
\end{eqnarray}
etc. 
We calculated the moments $\mu_m$ and $\mu^{(p,n)}_m$ up to $m=6$.
\medskip

\noindent {\bf Motivation:}
The eigenstate of the Rabi Hamiltonian with $g=0$ (\ref{b1}) was used as 
a trial function for the $t$-expansion in Ref. \cite{Fess02b}.
For the CMX extrapolation with 5 connected moments, the numerical results
for $E_0$ are satisfactory only in the region of small $g$.
Amore et al. \cite{Amo} used the CMX scheme with up to 99 connected moments,
without a real improvement of the previous results for intermediate and
large values of $g$; in some regions of model's parameters, they even
encounter numerical instabilities (considerable oscillations) of the results.
The authors conclude that the method is not reliable for practical purposes.
\medskip

\noindent {\bf The method:}
Our idea is to use the CMX and CSM versions of the $t$-expansion method
with the Rabi variational trial functions, the non-symmetrized 
$\vert \psi(x,y)\rangle$ (\ref{e7}) and the $(p)$, $(n)$ symmetrized 
$\vert \psi^{(p,n)}(x,y)\rangle$ (\ref{e10}).
Using the $t$-expansion with $m=1,3,4,\ldots$ connected moments involved, 
we have at disposal $E_0^{(m)}$ which depends on free parameters $\{ x,y \}$.
In the lowest (variational) order $m=1$, the optimized values of 
$\{ x_{\rm opt}^{(1)},y_{\rm opt}^{(1)} \} \equiv \{ x_{\rm opt},y_{\rm opt} \}$
are determined by the stationarity conditions (\ref{b2}) which imply the
global energy minimum in the $(x,y)$ space.
The determination of the free parameters for $m\ge 3$ is based on 
the following arguments.
Although the limit $\lim_{m\to\infty} E_0^{(m)}(x,y)$, if it exists, would not 
depend on $x$ and $y$, our finite truncations $E_0^{(m)}$ do.
If the free parameters are chosen properly in a convergence range of 
the series, $E_0^{(m)}$ converge smoothly to the exact $E_0$ as $m\to\infty$. 
In the opposite case, $E_0^{(m)}$ oscillates quickly as $m$ increases 
which is an indication of loss of convergence properties.  
To ensure at least a ``local independence'' of $E_0^{(m)}$ on free parameters, 
we impose the stationarity conditions \cite{Kol}
\begin{equation} \label{e5}
\frac{\partial E_0^{(m)}}{\partial x}=\frac{\partial E_0^{(m)}}{\partial y}=0.
\end{equation}
This equation determines $\{ x_{\rm opt}^{(m)},y_{\rm opt}^{(m)} \}$. 
In contrast to the variational $E_0^{(1)}$, the optimized $E_0^{(m)}$ is not 
a rigorous upper bound for the ground state energy.
Sometimes, there exist more solutions of Eqs. (\ref{e5}).
It is obvious not to accept maxima and saddle points, but still we can
have several minima and now the global one need not to be the best choice.
In general, there exists only a unique curve (hypersurface) of optimized 
$x_{\rm opt}^{(m)}(\omega_0,\omega,g)$, and of the coupled
$y_{\rm opt}^{(m)}(\omega_0,\omega,g)$, which is continuous in Rabi's 
parameters and simultaneously lies close to the variational solution; 
this will be our physical solution.
All other non-physical solutions, forming disconnected ``blind arms'',
are ignored; details for specific cases will be given bellow.
\medskip

\noindent {\bf Numerical results:}
We apply both CMX method (\ref{e3}) in $m=1,3,5$ orders
and CSM method (\ref{e4}) in $m=1,3,4,5,6$ orders.
The difference between the CMX and CSM results turns out to be very small.
In overwhelming number of cases, the results are slightly above 
the best (``exact'') estimates obtained by the straightforward diagonalization
of the Hamiltonian matrix in an appropriate basis set \cite{Amo,Bish}.
In the Rabi Hamiltonian, one parameter can be fixed as it merely sets 
the energy scale; we prefer to set $\omega_0=1$. 
Then we choose some $\omega$ and gradually change $g$ in the whole
interval $[0,\infty]$.
Estimates of $E_0^{(m)}(\omega_0=1,\omega,g)$ are expected to be satisfactory
for small and large values of $g$ as the trial functions are, by construction, 
close to the exact solutions at $g=0$ and $g\to\infty$ 
($\omega_0=0$ previously).
The true problem is the region of intermediate values of $g$. 

\begin{figure} [t]
\begin{center}
\includegraphics[scale=0.34]{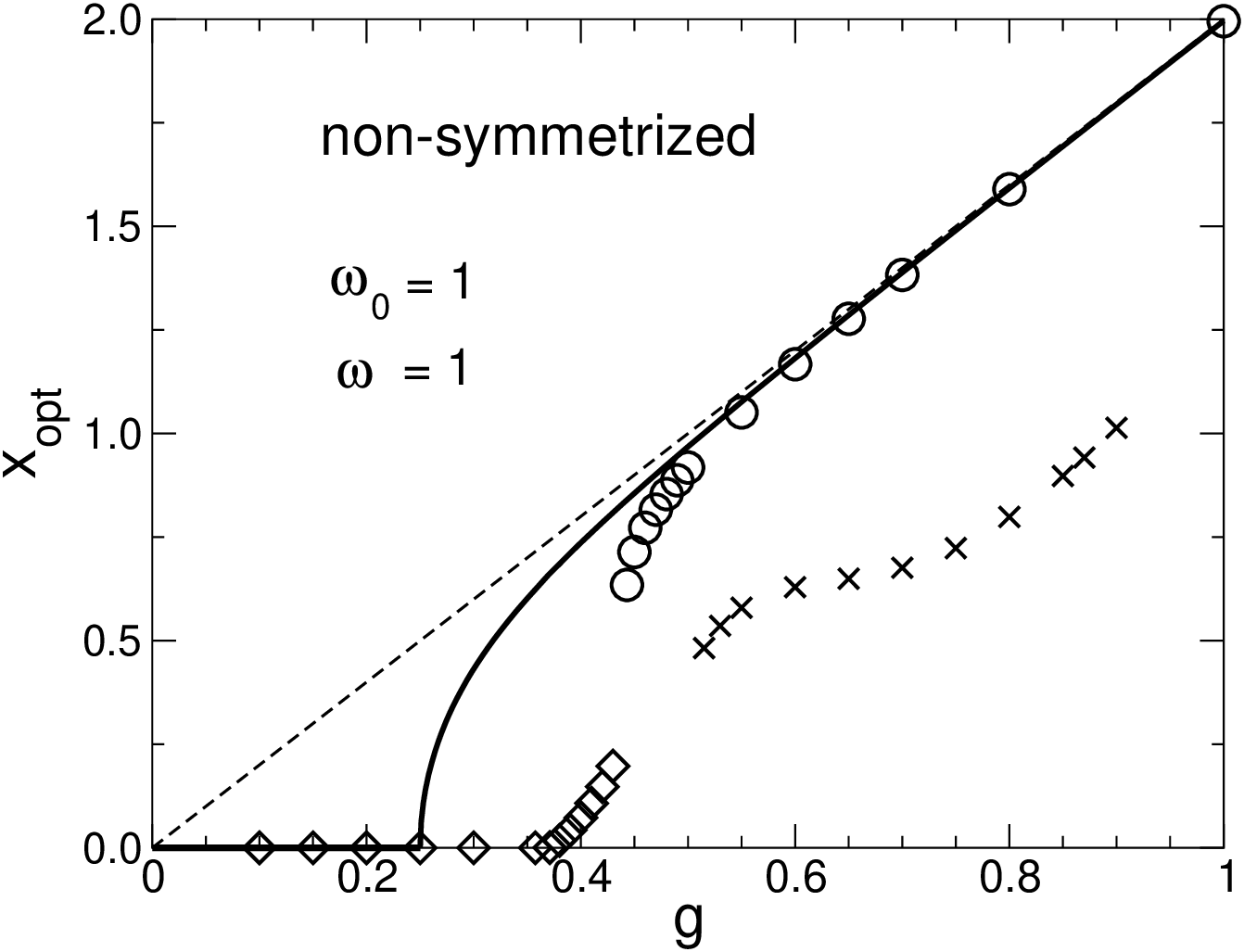}
\end{center}
\caption{The plot of $x_{\rm opt}(g)$ for $\omega_0=\omega=1$, calculated
with the non-symmetrized trial function (\ref{e7}). 
The solid curve is the variational ground state energy $E_0^{(1)}$, 
the dashed line is the large $g$-asymptotic. 
The CSM estimate $E_0^{(6)}$ has three blind arms denoted by symbols 
$\bigcirc$, $\diamondsuit$ and $\times$.}
\label{fig1}
\end{figure}

First we present the results for the non-symmetrized trial function 
$\vert \psi(x,y)\rangle$ (\ref{e7}). 
For any $\{ \omega_0,\omega,g \}$, the variational $E_0^{(1)}$ possesses just
one minimum at $\{x_{\rm opt}, y_{\rm opt}\}$ lying in the interval (\ref{e9}). 
Each of the functions $x_{\rm opt}(g)$ and $y_{\rm opt}(g)$ is continuous.
The plot of $x_{\rm opt}(g)$ is represented in Fig. \ref{fig1} by 
the solid curve [the picture is similar for $y_{opt}(g)$]. 
For $16 g^2 \le \omega_0\omega$, the solution of Eq. (\ref{e5}) giving 
the minimum of $E_0^{(1)}$ is trivial: $\{x_{\rm opt}, y_{\rm opt}\}=\{ 0,0 \}$.
This is why the curve $x_{\rm opt}(g)$ lies on the $g$ axis up to $g=1/4$. 
If $g>1/4$, the solution becomes non-trivial and approaches the asymptotic 
$x_{\rm opt} = 2g/\omega$ (dashed line) for large $g$.
Although this curve is continuous, it is non-analytic at the point $g=1/4$.
Going to $E_0^{(m)}$ with $m\ge3$, the number of minimum solutions 
to Eq. (\ref{e5}) can be larger than one in certain intervals of $g$.  
The minima curves can break at some points (beyond which there are no 
minimum solutions) or split into several curves. 
None of them goes continuously from small to large values of $g$.

Nevertheless, the curves denoted by open diamonds and circles in 
Fig. \ref{fig1} can be used for small-$g$ and large-$g$ cases, respectively,
to obtain satisfactory results.
For small $g$, the trivial minimum $\{0,0\}$ can be directly inserted 
into all expressions. 
The variational $E_0^{(1)}(0,0)=-\omega_0/2$ and 
\begin{equation} \label{e15}
E_0^{(3)}(0,0) = -\frac{\omega_0}{2} - \frac{4 g^2}{\omega+\omega_0} .
\end{equation}
For $g\ll \omega_0$, this formula is consistent with the exact result with 
the relative error $\vert E_0^{(3)}(0,0)-E_0^{\rm exact}\vert/E_0^{\rm exact}$
of the order $10^{-5}$. 
The large-$g$ results, illustrated in the first window of Table 1, 
are even more precise.
We present the variational result $E_0^{(1)}$, the CSM $E_0^{(6)}$ and 
the numerically exact result \cite{Amo}. 
The values of $\{ x_{\rm opt},y_{\rm opt} \}$ and 
$\{ x_{\rm opt}^{(6)},y_{\rm opt}^{(6)} \}$ are very close to 
the asymptotic result $\{ 2g/\omega,-1 \}$.
For example, for $\omega_0=\omega=1$ and $g=5$, the minimum of $E_0^{(6)}$ is 
at $\{ 9.99997, -0.997364 \}$.
The formulas with symmetrized trial functions (see bellow) lead to 
the results with comparable accuracy in both small-$g$ and large-$g$ regions.

\begin{figure} [t]
\begin{center}
\includegraphics[scale=0.34]{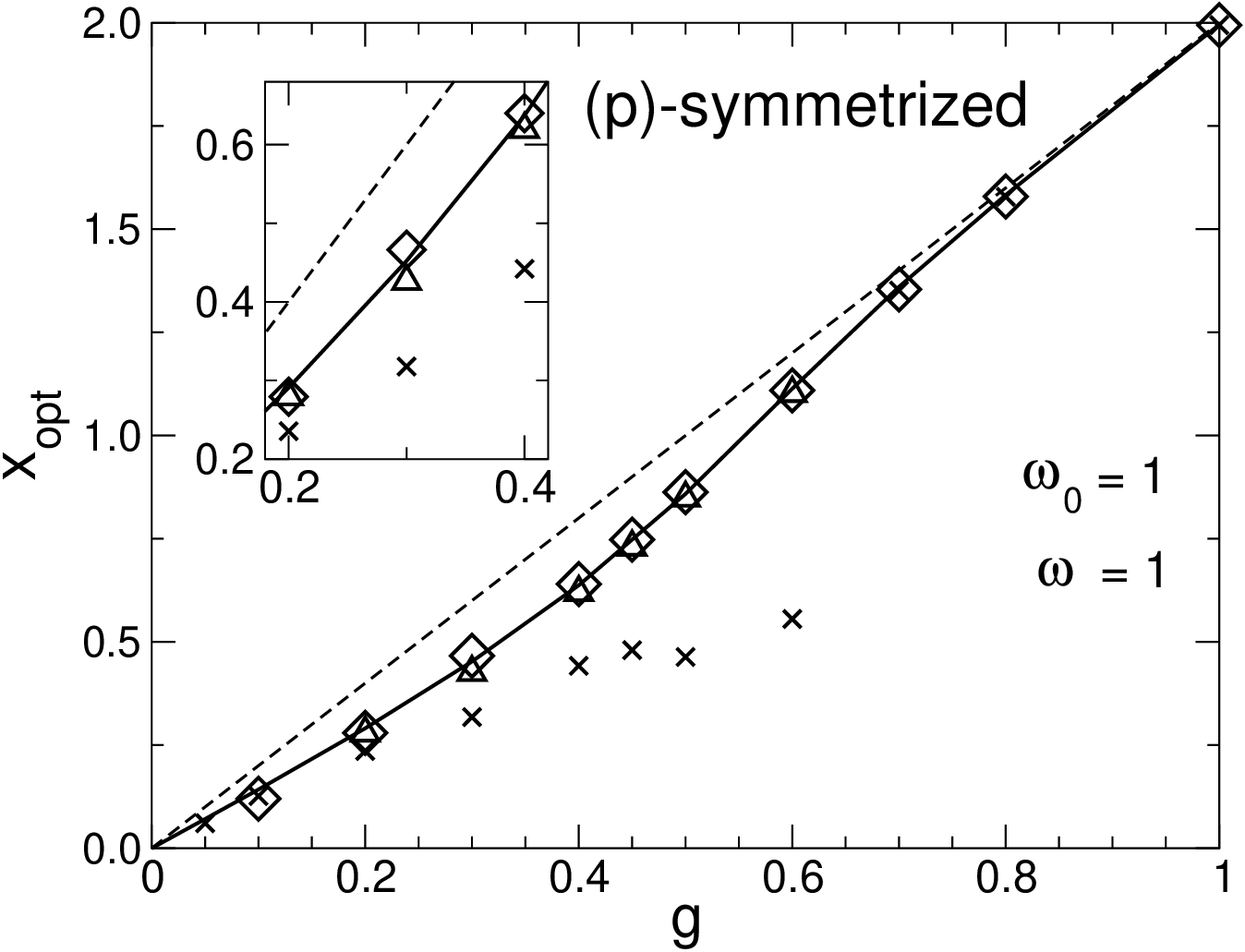}
\end{center}
\caption{The plot of $x_{\rm opt}(g)$ for $\omega_0=\omega=1$, calculated
with the $(p)$ symmetrized trial function (\ref{e10}).  
The solid curve corresponds to the variational $E_0^{(1,p)}$, 
the dashed line is the large $g$-asymptotic. 
The CMX estimate $E_0^{(5,p)}$: Optimized parameters closest to the solid
curve are denoted by $\diamondsuit$, 
blind arms are denoted by $\triangle$ and $\times$.}
\label{fig2}
\end{figure}

\begin{table}
{Table 1. Estimates of the energies $E_0$ and $E_1$ for the Rabi Hamiltonian
with $g=5$.}
\smallskip

{\begin{tabular}{cccccc}
 \hline
\ & $\omega_0=\omega=1$ & $\omega_0=1, \omega=2$ \\
 \hline
 \ $E_0^{(1)}$ & -100.006250000 & -50.001250000 \\
 \ $E_0^{(6)}$      & -100.006265682 & -50.001262703 \\
 \ $E_0^{\rm exact}$   & -100.006265704 & -50.001262758 \\
 \hline
 \ $E_1^{(1,n)}$ & -100.006250000 & -50.001250000 \\
 \ $E_1^{(6,n)}$      & -100.006265686 & -50.001262703 \\
 \hline
\end{tabular}}
\end{table}

The results for the ground state energy obtained with the $(p)$-symmetrized 
trial function $\vert \psi^{(p)}(x,y)\rangle$ (\ref{e10}) are presented
in Fig. \ref{fig2}. 
The main advantage of the variational result $E_0^{(1,p)}$ in comparison
with the non-symmetrized one is that the curve of unique minima, 
plotted as the solid curve, is not only continuous but also analytic for all 
$g$.
In the considered higher orders $E_0^{(m,p)}$ $(3\le m\le 6)$ and for both 
CSM and CMX methods, there exists a unique counterpart of the variational 
curve of minima (represented by open diamonds) which is continuous and 
free of singular points in the whole interval of $g$ values; 
these are the accepted optimized minima. 
Similarly as for the non-symmetrized trial function, blind disconnected 
curves of minima appear (open triangles and crosses); we ignore them. 
Since some of minima are very close to the variational curve,
the region of intermediate values of $g$ is magnified in the inset of 
Fig. \ref{fig2}. 

\begin{figure} [t]
\begin{center}
\includegraphics[scale=0.33]{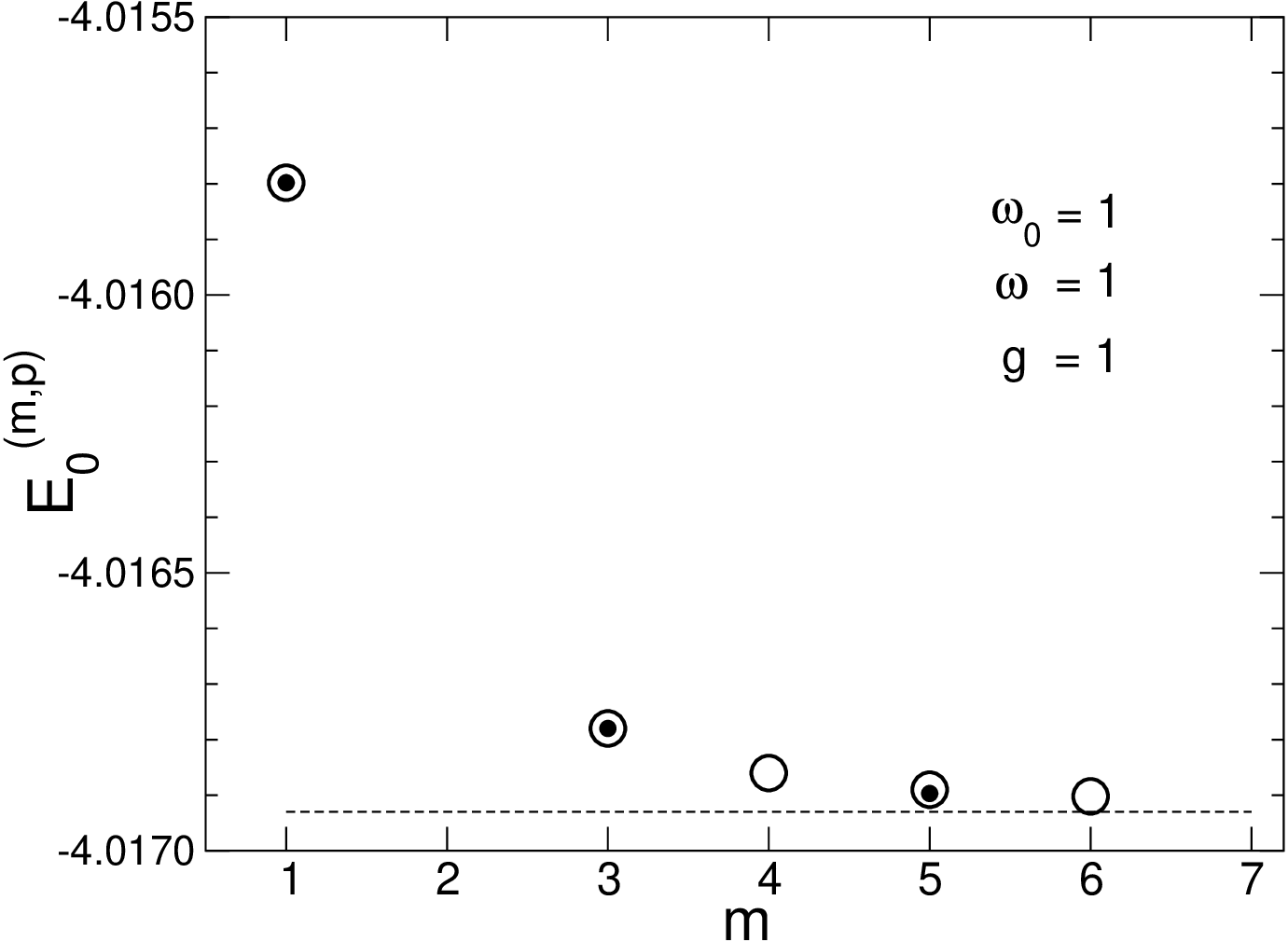}
\end{center}
\caption{The ground-state energy $E_0^{(m,p)}$ for $\omega_0=\omega=g=1$,
calculated with the $(p)$-symmetrized trial function:
The CSM results $\bigcirc$ in $m=1,3,4,5,6$ orders, the CMX results $\bullet$
in $m=1,3,5$ orders. 
Dashed line is the numerically exact value \cite{Bish}.}
\label{fig3}
\end{figure}

\begin{figure} [t]
\begin{center}
\includegraphics[scale=0.33]{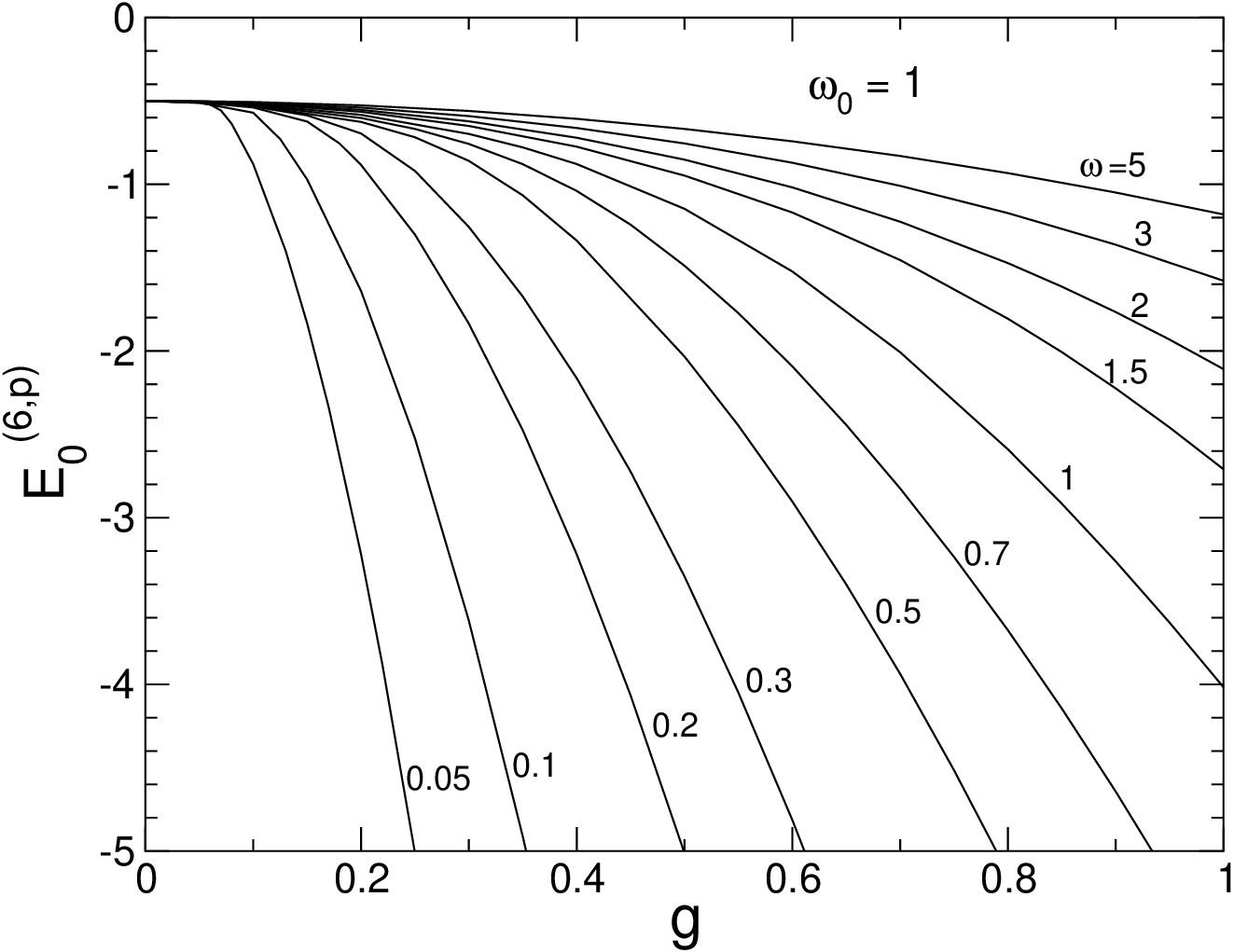}
\end{center}
\caption{The CSM estimate of the ground state energy with the $(p)$ 
symmetrized trial function, $E_0^{(6,p)}$, for $\omega_0=1$, various 
discrete values of $\omega$ and the interaction parameters $g$
constrained to the interval $[0,1]$.}
\label{fig4}
\end{figure}

First we discuss the results for small-$g$.
Bishop et al. \cite{Bish} showed that for $g\ll \omega_0$ the coordinates 
of minima for the variational $E_0^{(1,p)} = \mu_1^{(p)}$ 
[see Eq. (\ref{e14})] are, up to the term linear in $g$, given by
\begin{equation} \label{e16}
x_{\rm opt} \approx \frac{2g}{\sqrt{\omega(\omega_0+\omega)}} ,
\quad y_{\rm opt} \approx -\frac{2g}{\omega_0+\omega} .
\end{equation}
Inserting these values into $E_0^{(1,p)}$ and expanding in small $g$ up to 
the $g^2$ term we reproduce Eq. (\ref{e15}), derived from 
the non-symmetrized $E_0^{(3)}$.
This coincidence confirms that the non-symmetrized trial function gives 
adequate results not only for large $g$, where the corresponding formulas 
effectively merge because the corresponding $x$ is large, 
but also in the region of small $g$. 

For Rabi's parameters $\omega_0=\omega=g=1$, a quick convergence of 
the results for the ground state energy $E_0^{(m,p)}$ to the exact value as 
$m$ increases is shown in Fig. \ref{fig3}. 
The CMX data are represented by full circles, the CSM data by open circles;
note that the CMX and CSM results are very close to each other.
The (numerically) exact value \cite{Bish} is represented by the dashed line. 
The improvement of the variational $m=1$ result is remarkable already for
$m=3$.
As $m$ increases, the convergence of the data to the exact value is excellent. 

Our procedure enabled us to calculate very quickly the CSM ground state 
energies $E_0^{(6,p)}$ for various sets of the Rabi Hamiltonian parameters, 
see Fig. \ref{fig4}. 
Without any loss of generality we set $\omega_0=1$. 
Each curve is labeled by the boson energy $0.05\le \omega\le 5$. 
The interaction parameter $g$ is constrained to $0\le g\le 1$.
All values are correct within the resolution of the plots.
The worst relative error of the order $10^{-4}$ was achieved for medium 
values of $g$. 

\begin{figure} [t]
\begin{center}
\includegraphics[scale=0.33]{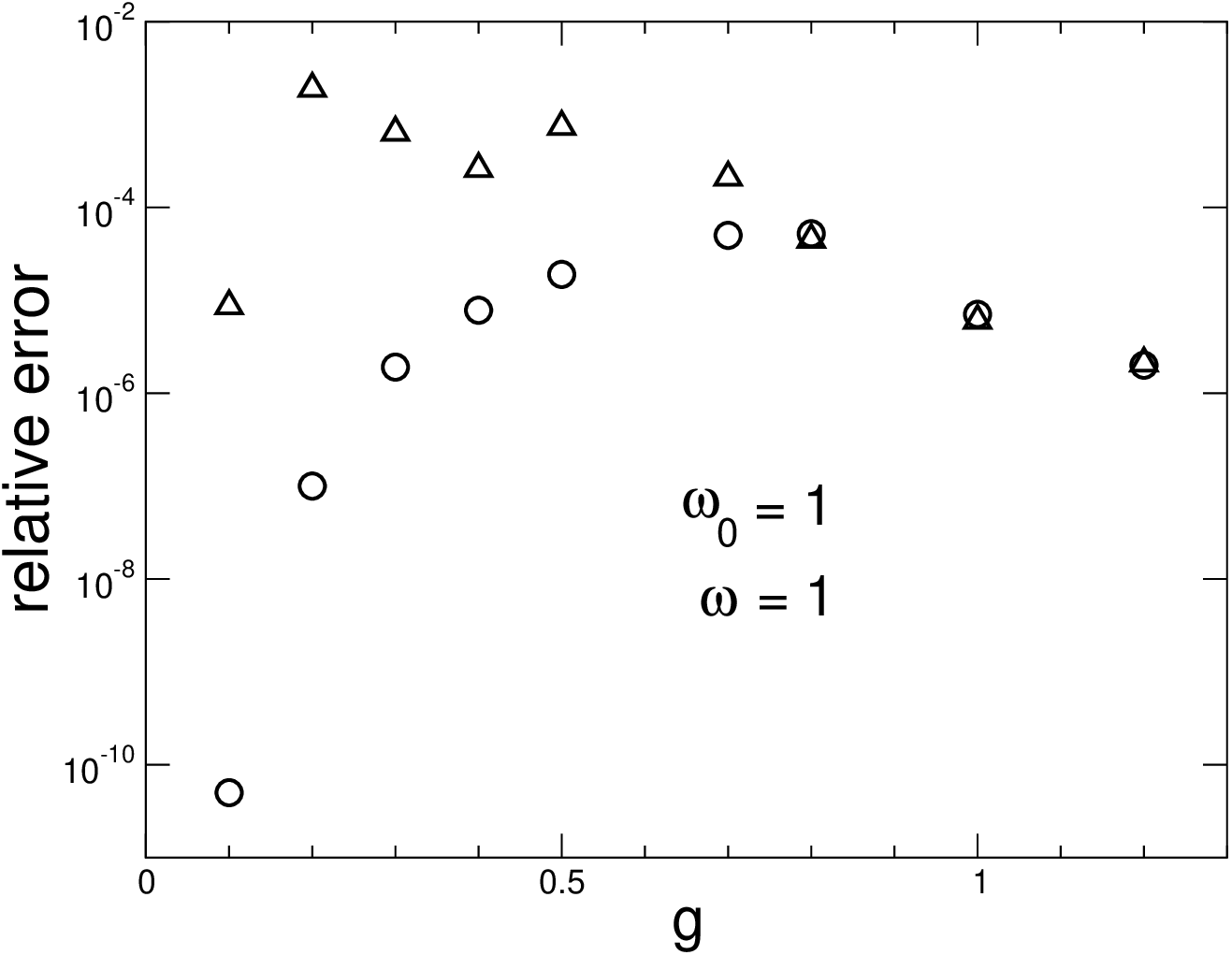}
\end{center}
\caption{The relative error of the CSM estimates of the ground state energy 
$E_0^{(6,p)}$ [$(p)$ symmetrized trial function, open circles] and the first 
excited energy $E_1^{(6,n)}$ [$(n)$ symmetrized trial function, open triangles], 
for $\omega_0=\omega_1=1$ and an interval of $g$-values.}
\label{fig5}
\end{figure}

To document an extraordinary accuracy of the obtained results,
in Fig. \ref{fig5} we present the relative error of the CSM estimates of
the ground state energy $E_0^{(6,p)}$ [$(p)$ symmetrized trial function, 
open circles] and of the first excited energy $E_1^{(6,n)}$ 
[$(n)$ symmetrized trial function, open triangles] for 
$\omega_0=\omega_1=1$ and an interval of $g$-values.
Within the whole parameter space of the Rabi Hamiltonian, the relative error 
is smaller than $10^{-4}$ for $E_0$ and smaller than $10^{-2}$ for $E_1$. 
Like for example, our CSM results for the ground state energy $E_0^{(6,p)}$
are $-0.69761396$ for $g=0.3$ and $-0.87854267$ for $g=0.4$, the (numerically) 
exact values $E_0^{\rm exact}$ are $-0.69761529$ for $g=0.3$ and  
$-0.87854932$ for $g=0.4$ \cite{Amo-pc}. 
The relative errors are $1.9\times 10^{-6}$ and $7.8\times 10^{-6}$, 
respectively.
As concerns the first excited energy, Bishop et al. \cite{Bish} reported 
the largest relative error (almost $0.4$) for the variational estimate 
$E_1^{(1,n)}$ at the interaction constant $g=0.2$. 
We see in Table 2 that this error goes down quickly in higher approximation 
orders.

\begin{table}
{Table 2. Estimates and relative errors of $E_1$ 
for $\omega_0=\omega=1$ and $g=0.2$. \phantom{aaaaaaaaaaaaaaaaa} }
\smallskip
{\begin{tabular}{cccccc}
 \hline
 \  & Estimate & Rel. error \\
 \hline
 \ $E_1^{(1,n)}$ var.& 0.00324806 & 0.39 \\
 \ $E_1^{(5,n)}$ CMX & 0.00233753 & 0.000335 \\
 \ $E_1^{(6,n)}$ CSM & 0.00234135 & 0.00197 \\
 \ $E_1^{\rm exact}$    & 0.00233675 & 0 \\
 \hline
\end{tabular}}
\end{table}
\medskip

\noindent {\bf Conclusion:}
In conclusion, it turns out that the optimization of low orders of the
$t$-expansion for the Rabi Hamiltonian improves remarkably the precision 
of the variational ground state estimates.
In the whole parameter range of Rabi model, with the symmetrized trial 
function the relative error is smaller than $10^{-4}$ ($0.01$\%). 
The accuracy of the $t$-expansion method is not ensured if the trial function 
is not properly chosen, as was seen in the case of non-symmetrized trial 
function and medium values of the interaction constant $g$. 

\section*{Acknowledgments}
We thank P. Amore for sending us high-precision values of $E_0$. 
This work was supported by the grants VEGA 2/0113/2010 and CE-SAS QUTE.

\end{document}